\documentclass[conference]{IEEEtran}
\IEEEoverridecommandlockouts
\usepackage{cite}
\usepackage{amsmath,amssymb,amsfonts}
\usepackage{algorithmic}
\usepackage{graphicx}
\graphicspath{{Figures/}}
\usepackage{subfigure}
\usepackage{textcomp}
\usepackage{xcolor}
\usepackage{multirow}
\def\BibTeX{{\rm B\kern-.05em{\sc i\kern-.025em b}\kern-.08em
    T\kern-.1667em\lower.7ex\hbox{E}\kern-.125emX}}
\begin{document}

\title{Uplink Channel Impulse Response Based Secondary Carrier Prediction
}
\author{Prayag~Gowgi* and Vijaya Yajnanarayana\\
  Ericsson Research\\
  Email: *prayag.gowgi@ericsson.com}


\maketitle

\begin{abstract}
A typical handover problem requires sequence of complex signaling between a UE, the serving, and target base station. In many handover problems the down link based measurements are transferred from a user equipment to a serving base station and the decision on handover is made on these measurements. These measurements together with the signaling between the user equipment and the serving base station is computationally expensive and can potentially drain user equipment battery. Coupled with this, the future networks are densely deployed with multiple frequency layers, rendering current handover mechanisms sub-optimal, necessitating newer methods that can improve energy efficiency. In this study, we will investigate a ML based approach towards secondary carrier prediction for inter-frequency handover using the up-link reference signals. 
\end{abstract}


\section{Introduction}

\begin{figure}
\centering
\includegraphics[scale=0.22]{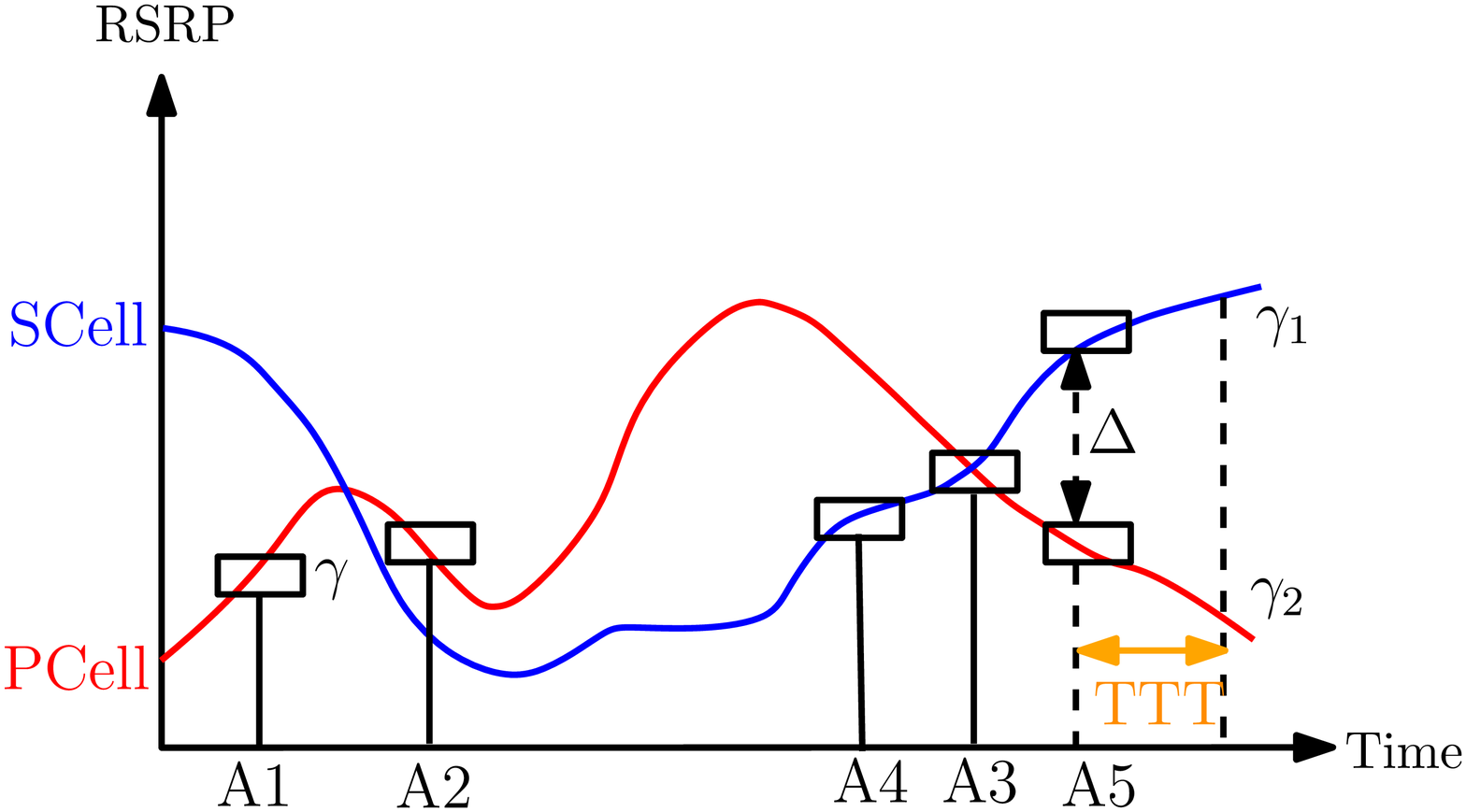}
\caption{Various events in a typical HO procedure: RSRP of both PCell and SCell are plotted against time. Event A1 is triggered when RSRP of the primary cell (PCell) is above certain threshold $\gamma>0$, event A2 is triggered when RSRP of PCell is below certain threshold, event A3 is triggered when RSRP of secondary cell (SCell) is above RSRP of PCell by an offset, event A4 is triggered when RSRP of SCell is above certain threshold, and event A5 is triggered when RSRPs of PCell is below a threshold $\gamma_{1}>0$ and SCell is above a threshold $\gamma_{2}>0$. The HO is triggered at the UE when the event A5 holds true for certain period of time called time-to-trigger (TTT).}
\label{Fig:Events}
\end{figure}

The handover (HO) of a user equipment (UE) is an integral part of a  multi-cell wireless communication system. With large mobility of UEs, HO becomes a challenging task in dense deployments. Recently, third generation partnership project (3GPP) has standardized the HO procedure for long term evolution (LTE) \cite{3GPP}. A UE periodically measures the signal on downlink (DL) from both serving and the target cell such as referenced symbols received power (RSRP) and the referenced symbols received quality (RSRQ). Whenever certain network configuration conditions are satisfied, the UE will send the latest measurement reports to the serving cell. These predefined network configuration conditions, called events, are set by the network through radio resource control configuration (RRC) messages. A typical RRC message consists of list of parameters to be measured, reporting criteria, and reporting configuration. Some of the important events relevant to HO are A1-A5 as shown in the Figure \ref{Fig:Events} and is summarized in Table \ref{tab:EventsTab}. A typical HO procedure with various events in action is shown in Figure \ref{Fig:Events} and the HO is triggered at the UE when the event A5 holds true for certain period of time called time-to-trigger (TTT) which makes the current approach \textit{reactive} to the events with each of the signaling between UE and the primary and the target cells are battery consuming which should be avoided as much as possible \cite{Onireti2015energy}.
\begin{table}[h]
\centering
\caption{Important events in a RRC configuration message. PCell: primary cell or the serving cell, SCell: target cell or the secondary cell, $\gamma>0$, $\delta>0$, $\gamma_{1}>0$, and $\gamma_{2}>0$ are thresholds. }
\label{tab:EventsTab}
\begin{tabular}{ll}
Events & Triggered when \\
A1 & PCell $\geq \gamma$ \\
A2 & PCell $< \gamma$ \\
A3 & SCell $\geq $ PCell$+\delta$\\
A4 & SCell $\geq \gamma$ \\
A5 & SCell $\geq \gamma_{1}$ and PCell $\leq \gamma_{2}$
\end{tabular}
\end{table}


\begin{figure}[ht]
    \centering
    \includegraphics[scale=0.25]{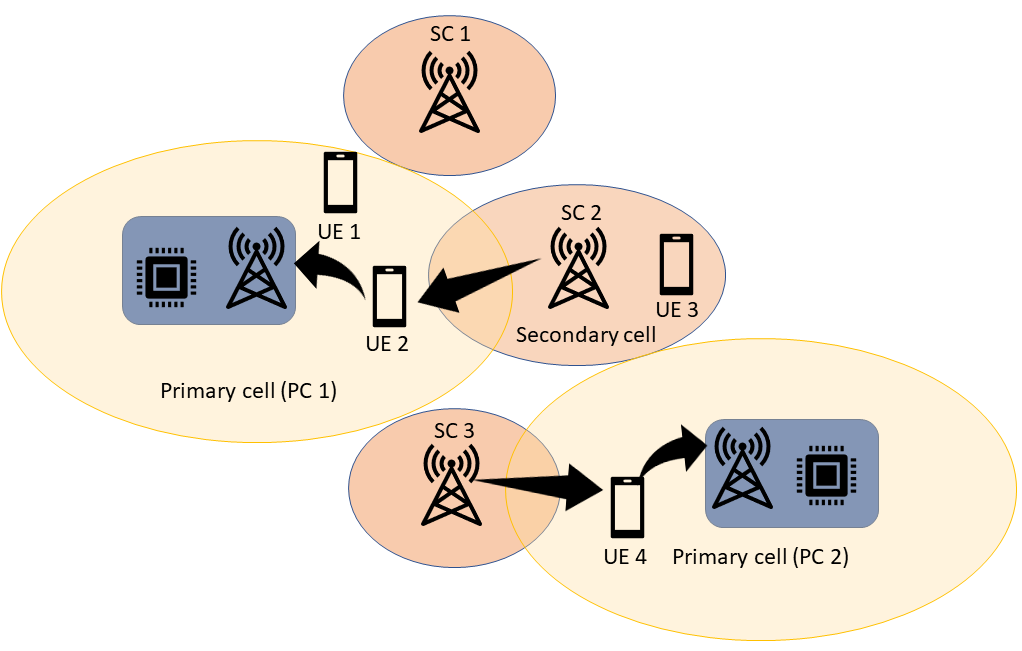}
    \caption{A scenario with four UEs, three secondary, and two primary cells. UE-1 and 2 are currently being served by primary cell (PCell)-1, UE-4 is served by PCell-2, and UE-3 is served by secondary cell (SCell)-2. UE-2 and 4 will measure RSRP levels from both PCell-1 and 2 SCell-2 and 3 and then PCell-1 and 2 will evaluate these measurement reports to make a decision on the HO.}
    \label{fig:IntroductionFigure}
\end{figure}

\begin{figure}[ht]
	\centering
	\includegraphics[scale=0.25]{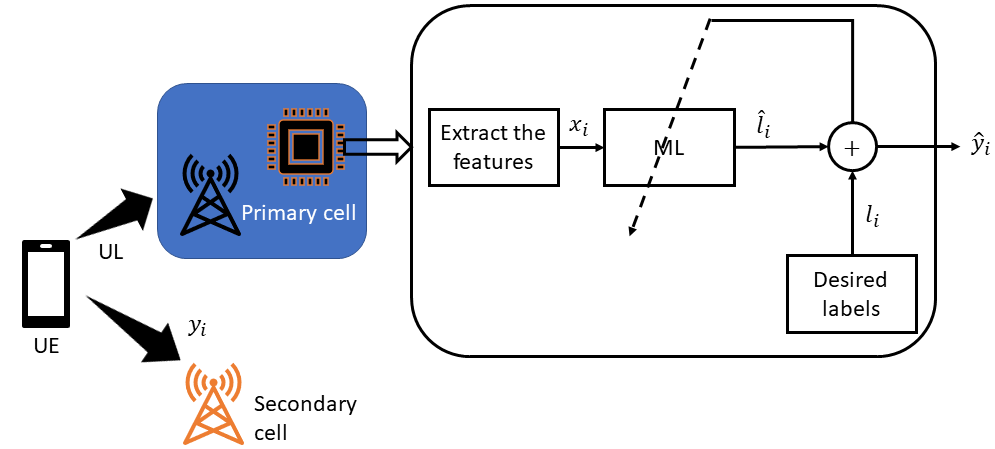}
	\caption{The general idea of our approach is as follows: a PCell currently serving a UE is equipped with a ML algorithm to make HO decision on the inter-frequency HO of the UE. The input for the ML algorithm is the relevant features extracted from the UL channel impulse response to the PCell. Upon completion of training of the ML algorithm, the existence or the non-existence of the SCell link is predicted by the ML algorithm at the PCell. }
	\label{fig:ScpArchitectureFigure}
\end{figure}
The future wireless network deployments in 5G and beyond are dense, they operate at multiple frequency bands, and support higher capacity by opportunistically selecting among multiple frequency bands. If the existing reactive system for the inter-frequency HO is used in such a deployment, then UE has to make excessive inter-frequency measurements owing to the fact that deployment is dense and operating at multiple frequencies. Upon measuring the signal levels the UE has to send the measurement report back to the serving cell, leading to increased energy consumption. Figure \ref{fig:IntroductionFigure} is a scenario with four UEs, three secondary, and two primary cells. UE-1 and UE-2 are currently being served by PCell-1, UE-4 is served by PCell-2, and UE-3 is served by SCell-2. UE-2 and UE-4 will measure RSRP levels from both PCell-1, PCell-2, SCell-2 and SCell-3 respectively. Now PCell-1 and PCell-2 will evaluate these measurement reports to make a decision on the HO. Machine learning (ML) techniques are suitable in such scenarios where the complex decision making tasks are driven by data. 

The advantages of ML based approach is that the HO decisions does not require inter-frequency measurements, reporting of the measurements back to the serving cell, simplifies the HO signal flow, greatly reduces the energy requirement by the UE, and simplifies the transceiver design of the UE. In addition, it is advantageous if there is a little or no modification to the existing standards. Therefore, we would like to utilize the existing UL reference signals that are typically available at the 5G antenna systems for beamforming decisions for making HO decisions which are lacking in some of the recent prior works \cite{Vj20195g,Henrik2018predicting,Ali,Feltrin}. Therefore, we investigate a ML based approach for inter-frequency HO using the UL reference signals.


\begin{figure}
	\centering
	\includegraphics[width=\linewidth]{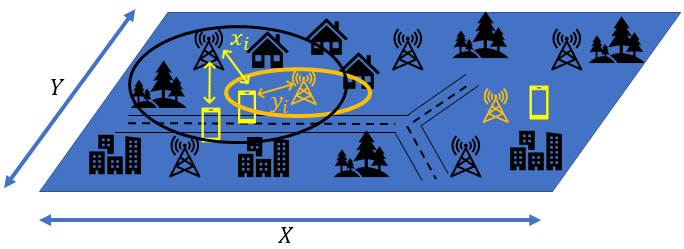}
	\caption{A general experimental setup. In a bounded geographical region i.e., $ X , Y \in [a \times b] \times [c \times d] $ where $ a,b,c,d \in \mathbb{R} $. For every UE whose UL link gain is above a threshold its UL impulse response is recorded at the PCell base station and certain features are extracted ($ x_{f_{p}} $) and for every such UE, we set label $ y=1 $ if $ G\left(x_{f_{s}}\right)\geq\alpha $ and $ 0 $ otherwise where $ G: \mathbb{R}^{d}\rightarrow\mathbb{R} $ and $\alpha>0$. }
	\label{Fig:GeneralScenario}
\end{figure}

\begin{figure*}[ht]
	\centering
	\minipage{0.32\textwidth}
		\includegraphics[width=\textwidth,height=0.7\linewidth]{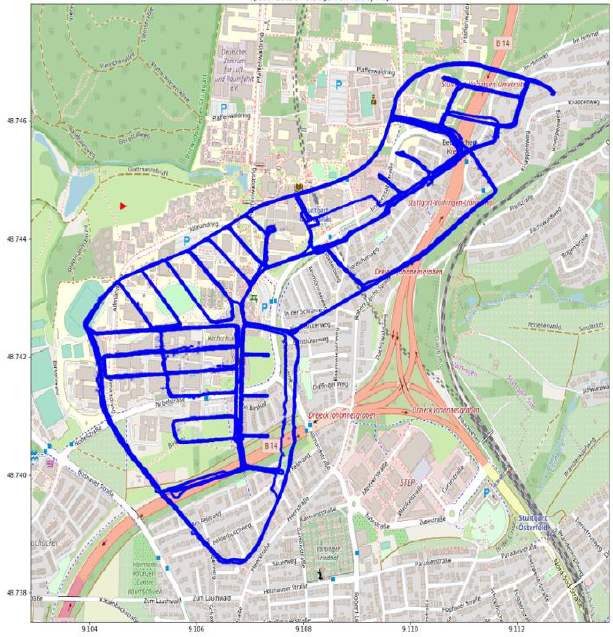}
		\caption{The path traversed by a moving vehicle with the transmitter.}
		\label{fig:D1MapWithPath}
	\endminipage
	\hspace{5cm}
	\minipage{0.32\textwidth}
		\includegraphics[width=\textwidth,height=0.7\linewidth]{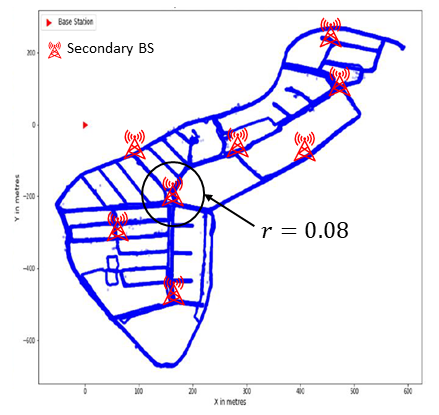}
		\caption{Addition of new secondary BSs: we assume the existence of secondary BSs at few junction points.}
		\label{fig:D1PathWithSecCells}
	\endminipage\hfill
\end{figure*}
\begin{figure*}
	\centering
		\begin{minipage}[b]{0.3\textwidth}
		\includegraphics[width=\textwidth]{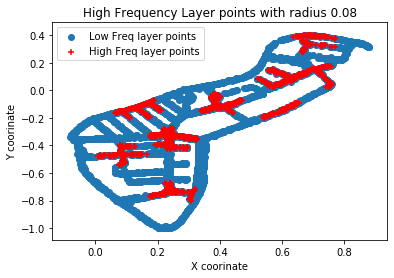}
			\caption{Label generation: consider all the UEs within a circle of radius equal to 0.08 to have a link with the newly added BSs and label them as "1" ($\color{red}+$), otherwise as "0" ($\color{blue}\bullet$). }
		\label{fig:R0p08}
	\end{minipage}
	\hspace{5cm}
	\begin{minipage}[b]{0.3\textwidth}
	\includegraphics[width=\textwidth]{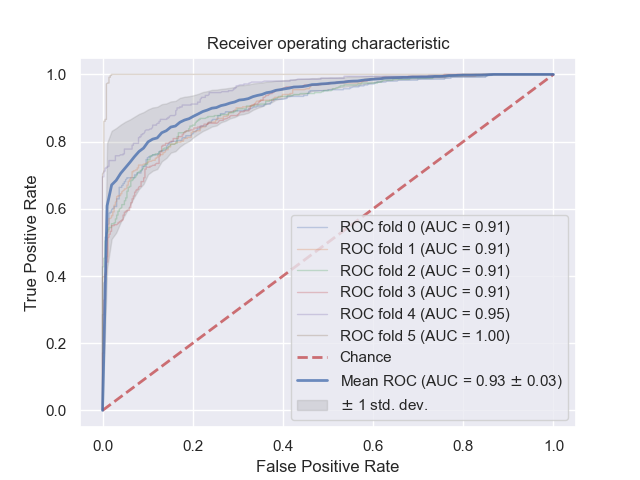}
	\caption{ROC for the scenario D1: binary classification is performed using 6-fold cross validation on the hybrid sampled data set.}
	\label{fig:IEEEDataROC}
	\end{minipage}
	\begin{minipage}[b]{0.3\textwidth}
	\includegraphics[width=\textwidth]{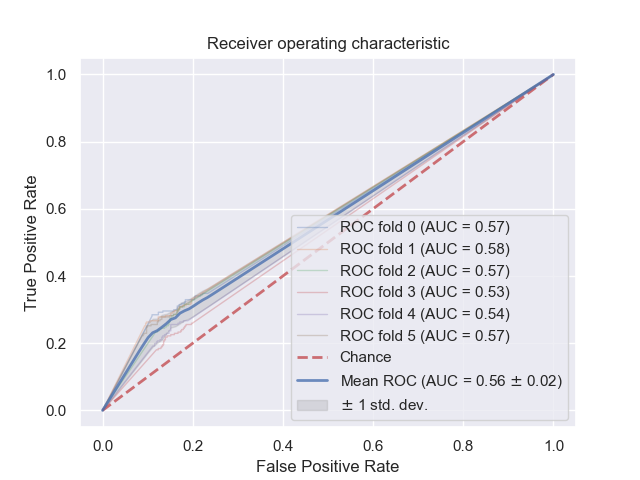}
	\caption{ROC for the scenario D1 (imbalanced) using MAP approach with a Gaussian distribution for the likelihood.}
	\label{fig:MAPROC}
	\end{minipage}
	\hspace{5cm}
	\begin{minipage}[b]{0.3\textwidth}
		\includegraphics[width=\textwidth]{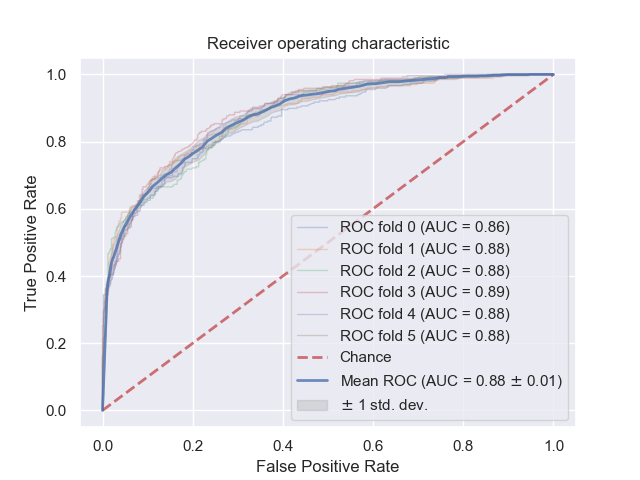}
		\caption{Comparison of ROC for the scenario D1 (imbalanced) with MAP approach.}
		\label{fig:D1ROCImbalancedData}
	\end{minipage}
\end{figure*}

\begin{figure*}[ht]
	\centering
	\begin{minipage}[b]{0.3\textwidth}
		\includegraphics[width=\textwidth]{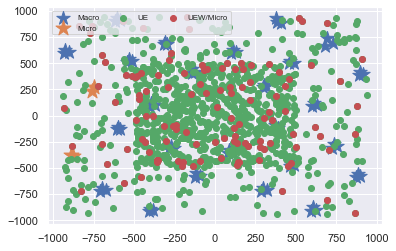}
		\caption{Scenario D2.a: 75 macro BSs and 5 micro BSs}
		\label{fig:AsianCityScenario80BS1000UE5MiBs}
	\end{minipage}
	\hspace{5cm}
	\begin{minipage}[b]{0.3\textwidth}
		\includegraphics[width=\textwidth]{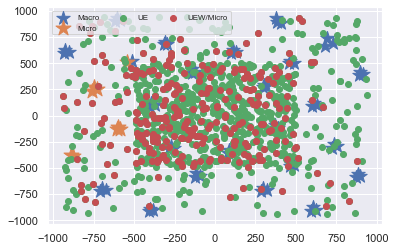}
		\caption{Scenario D2.b: 70 macro BSs and 10 micro BSs}
		\label{fig:AsianCityScenario80BS1000UE10MiBs}
	\end{minipage}
	
	\begin{minipage}[b]{0.3\textwidth}
		\includegraphics[width=\textwidth]{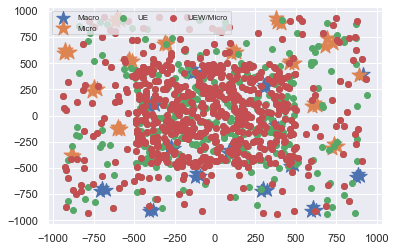}
		\caption{Scenario D2.c: 40 macro BSs and 40 micro BSs}
		\label{fig:AsianCityScenario80BS1000UE40MiBs}
	\end{minipage}
	\hspace{5cm}
	\begin{minipage}[b]{0.3\textwidth}
		\includegraphics[width=\textwidth]{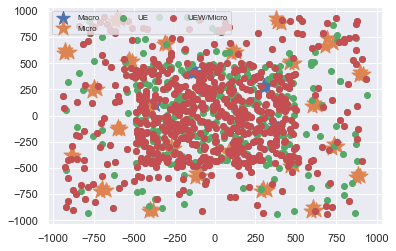}
		\caption{Scenario D2.d: 10 macro BSs and 70 micro BSs}
		\label{fig:AsianCityScenario80BS1000UE70MiB}
	\end{minipage}

  \begin{minipage}[b]{0.29\textwidth}
    \includegraphics[width=1.1\textwidth,height=0.75\textwidth]{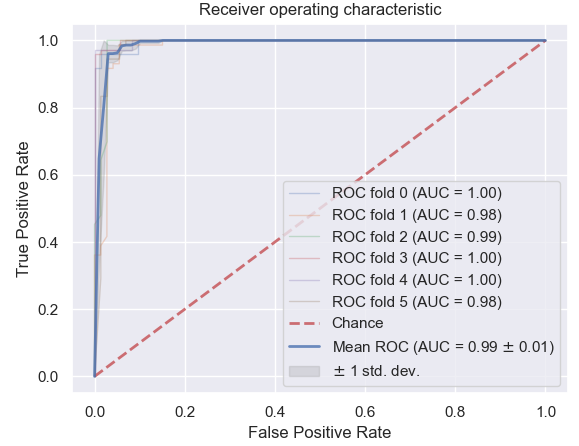}
    \caption{ROC for the scenario D2.a}
    \label{fig:AsianCityMic05}
  \end{minipage}
  \hspace{5cm}
  \begin{minipage}[b]{0.29\textwidth}
    \includegraphics[width=1.1\textwidth,height=0.75\textwidth]{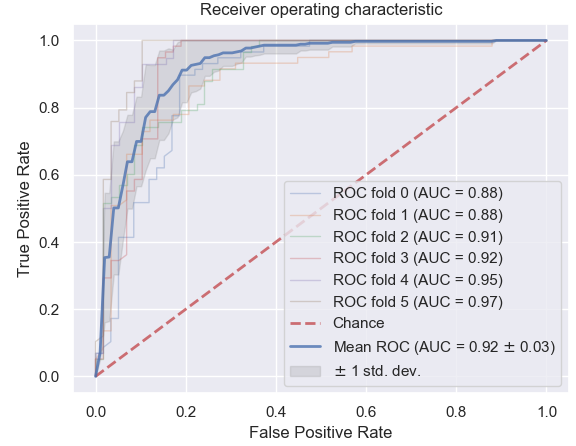}
    \caption{ROC for the scenario D2.b}
    \label{fig:AsianCityMic10}
  \end{minipage}
  
  \begin{minipage}[b]{0.29\textwidth}
    \includegraphics[width=1.2\textwidth,height=0.8\textwidth]{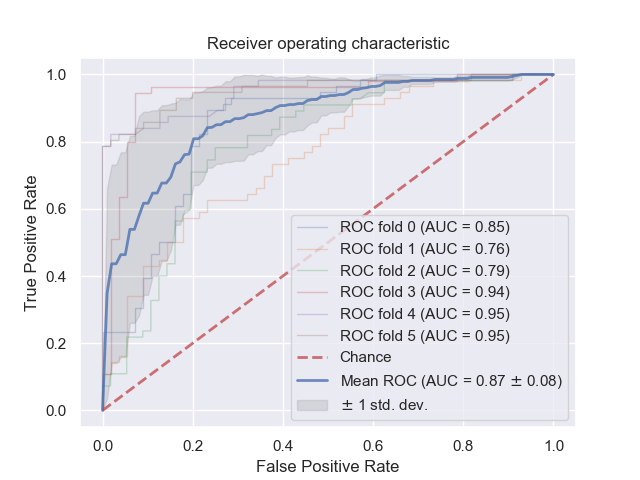}
    \caption{ROC for the scenario D2.c}
    \label{fig:AsianCityMic40}
  \end{minipage}
  \hspace{5cm}
  \begin{minipage}[b]{0.29\textwidth}
    \includegraphics[width=1.2\textwidth,height=0.8\textwidth]{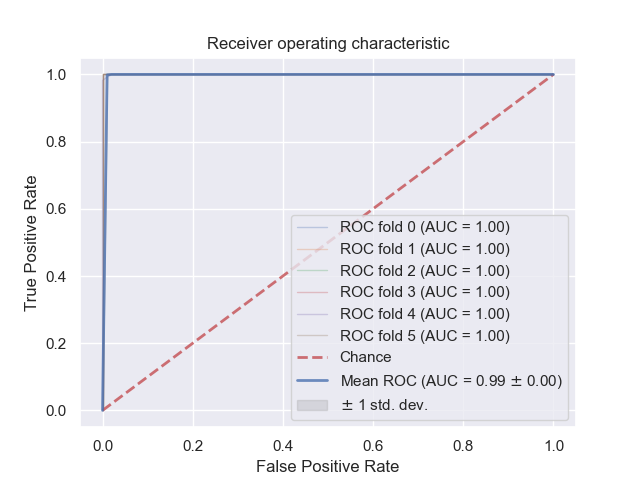}
    \caption{ROC for the scenario D2.d}
    \label{fig:AsianCityMic70}
  \end{minipage}
\end{figure*}

\begin{figure*}[ht]
	\centering
	\minipage{0.32\textwidth}
		\includegraphics[width=\linewidth]{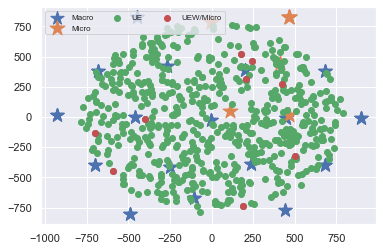}
		\caption{Scenario D3.a: 50 macro BSs and 10 micro BSs}
		\label{fig:Scenario600UE50Ma10MiBs}
	\endminipage\hfill
	\minipage{0.32\textwidth}
		\includegraphics[width=\linewidth]{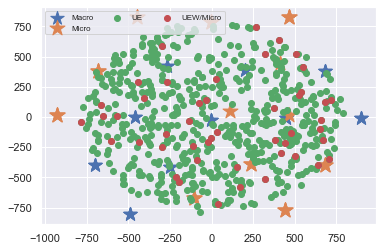}
		\caption{Scenario D3.b: 30 macro BSs and 30 micro BSs}
		\label{fig:Scenario600UE30Ma29MiBs}
	\endminipage\hfill
	\minipage{0.32\textwidth}
		\includegraphics[width=\linewidth]{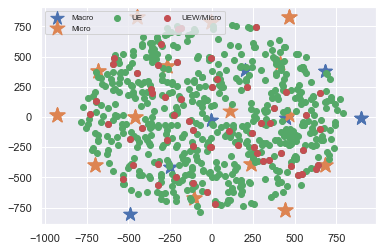}
		\caption{Scenario D3.c: 20 macro BSs and 40 micro BSs}
		\label{fig:Scenario600UE20Ma40MiBs}
	\endminipage\hfill
\end{figure*}

\begin{figure*}[ht]
  \centering
  \minipage{0.32\textwidth}
    \includegraphics[width=\linewidth]{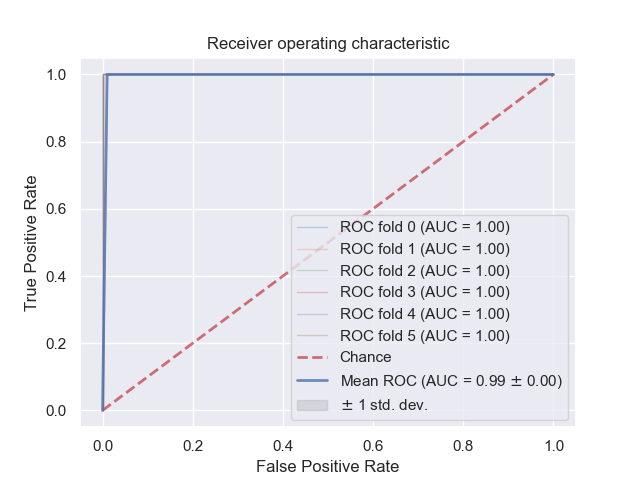}
    \caption{ROC for the scenario D3.a}
    \label{fig:LondonMi10}
  \endminipage\hfill
  \minipage{0.32\textwidth}
   \includegraphics[width=\linewidth]{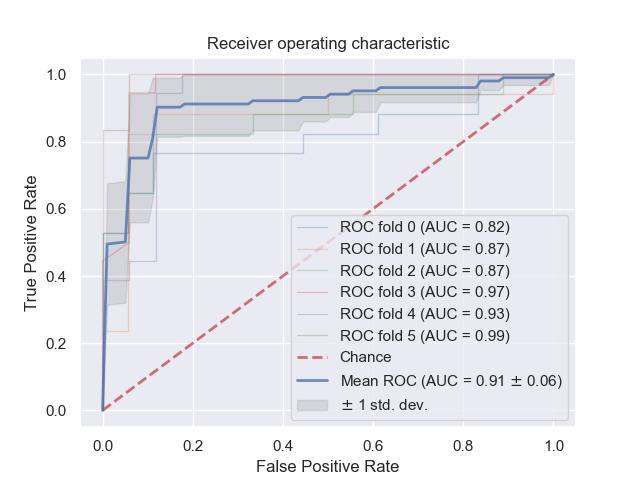}
    \caption{ROC for the scenario D3.b}
    \label{fig:LondonMi20}
\endminipage\hfill
\minipage{0.32\textwidth}
    \includegraphics[width=\linewidth]{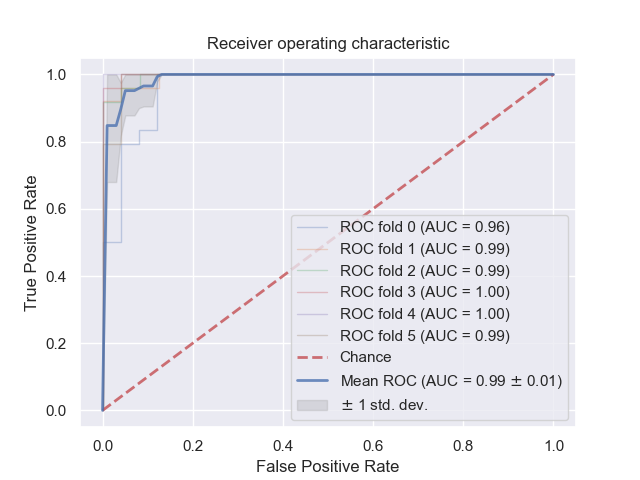}
    \caption{ROC for the scenario D3.c}
    \label{fig:LondonMi30}
    \endminipage\hfill
  \end{figure*}

The organization of the paper is as follows: in Section \ref{Sec:PriorWork} we discuss recent prior work in the context of our work, a maximum aposteriori probability decision rule is analyzed in Section \ref{Sec:Analysis}, and experimental setup along with the results are discussed in Section \ref{Sec:ExperimentalSetupAndResults} followed by conclusions in Section \ref{Sec:Conclusions}. 
	

\section{Prior work}
\label{Sec:PriorWork}
ML for HO problem is not new and in this section, we discuss some of the most recent works related to our work.
In \cite{Vj20195g}, authors propose a reinforcement learning based approach for HO control problem using UE measurement-report to maximize a long-term utility. The authors were able to successfully show the working of the approach by using average link-beam gain as utility for different propagation environment and mobility pattern. However, the proposed method does not address the inter-frequency HO problem. In \cite{Henrik2018predicting}, authors propose a ML based approach for handling inter-frequency HO problem. The goal of this technique is to predict the strongest cell on the secondary carrier based on primary carrier data. In this approach, authors looked at training an ML algorithm placed in the primary carrier cell to predict the strongest secondary carrier cell based on one or more of the following features: 1) RSRP of the serving and up to eight neighboring cells 2) precoder matrix indicator 3) timing advance (TA). Authors were able to successfully show the working of this approach however, the features considered are based on DL based-measurements by the UE. In this method, UE still have to do DL-measurements and report them to the BS resulting in large signaling overhead. Similar to \cite{Henrik2018predicting}, authors in \cite{Feltrin} proposed a neural network (NN) based approach to anticipate the HO and blind spots over a Wi-Fi network. In this approach NN considers the past M samples of received signal strength indicator (RSSI) and predicts HO and blind spots for the UE. RSSI is measured at the UE leading to faster battery draining of a UE. 
 Motivated by the idea of \cite{Henrik2018predicting} and considering the advantages that is discussed in the previous section, we propose a ML approach for predicting the existence or non-existence of secondary carrier coverage based on the already available UL reference signals on primary carrier. 

\section{System model}
\label{Sec:Architecture}
In this section we describe the system aspects, shown in Figure \ref{fig:ScpArchitectureFigure}, of our approach. For simplicity we consider the case of single UE, PCell, and SCell. The UE is currently being served by the PCell and the 5G antenna system at the PCell base station is equipped with a ML algorithm. In 5G and beyond systems the UL-impulse response are anyway available at the BS, as they are  used for beam-forming decisions. The ML algorithm of the PCell base station predicts the existence or non-existence of secondary carrier link for the UE. The existence or non-existence of a link is based on the signal strength  of the secondary carrier. In this work, we pre-define a threshold, and consider the existence of the secondary carrier if the signal strength is above this threshold. This formulation will aid us in treating this problem as a binary hypothesis problem where the output label "0" indicates "null" hypothesis (no secondary carrier detected) and "1" indicates the alternative hypothesis (presence of secondary carrier). The ground truth for training the ML algorithm is obtained by actual measurements.

\section{Analysis}
\label{Sec:Analysis}
In order to analyse the accuracy of the hypothesis we can resort to a maximum aposteriori probability (MAP) estimator whose decision criteria is discussed below.

Let $C$ and $x$ be the class label and observed data respectively, $P\left(C|x\right)$ be the posterior probability, $P\left(x|C\right)$ be the likelihood of $C$, and $P\left(C=1\right)=p$ and $P\left(C=0\right)=1-p$ be the prior probabilities. Assume that likelihood of $C$ follows a Gaussian distributions $\mathcal{N}(\mu_{0},\sigma^{2})$ and $\mathcal{N}(\mu_{1},\sigma^{2})$ for $ C=0 $ and $ C=1 $ respectively.  Following maximum a posteriori probability (MAP), we choose class one if 
                   \begin{align}
                        P\left(C=1|x\right) & \geq P\left(C=0|x\right) \\
                     \exp\left(\frac{\left(x-\mu_{0}\right)^{2}}{2\sigma^{2}}\right) p& \geq \exp\left(\frac{\left(x-\mu_{1}\right)^{2}}{2\sigma^{2}}\right) \left(1-p\right)\label{eqn:BayesianwithGaussian}.
                    \end{align}
Simplifying equation \eqref{eqn:BayesianwithGaussian} further, we get
\begin{equation}
    x \geq \frac{\sigma^{2}}{\left(\mu_{0}-\mu_{1}\right)}\log\left(\frac{1-p}{p}\right)+ \left(\frac{\mu_{0}+\mu_{1}}{2}\right)
    \label{eqn:Bayes}
\end{equation}
In many practical environments, the distributions $ P\left(x|C=1\right) $ and $ P\left(x|C=0\right) $ along with the prior probabilities are not known, thus AI based approaches are more favorable in those environments. 
\section{Experimental setup and results}
\label{Sec:ExperimentalSetupAndResults}

In this section, we describe the general experimental setup and how the simulation data was captured. We consider a bounded geographical region with large number of UEs, PCells, and SCells as shown in Figure \ref{Fig:GeneralScenario}. PCell is operating at a carrier frequency $ f_{p} $ and SCell is operating at a carrier frequency of $ f_{s} $. Typically $ f_{p}\in \mathrm{FR}_{1} $ and $ f_{s}\in \mathrm{FR}_{2} $\footnote{$ \mathrm{FR}_{1} $: 410-7125 MHz and $ \mathrm{FR}_{2} $: 24250-52600 MHz}, however, it is possible that both $ f_{p},f_{s}\in \mathrm{FR}_{1} $. For every UE whose UL link gain at PCell ($ f_{p} $) is above a particular threshold, the UL channel impulse response is recorded at the PCell base station and certain features are extracted ($ x_{f_{p}} \in \mathbb{R}^{d}$). Similarly for every such UE, we set $ y=1 $ if $ G\left(x_{f_{s}}\right)\geq\alpha $ and $ 0 $ otherwise
where $ G: \mathbb{R}^{d}\rightarrow\mathbb{R} $, $\alpha>0$, and the set of all $ \left(x_{i},y_{i}\right) $ forms the data set.

Following the procedure just described, we generate three data-sets:
data-set (D1) is derived from real world deployment, second, and third data-sets (D2 and D3) are derived from simulated datasets.

\begin{table}[ht]
	\caption{Three data sets, different scenarios (MB: \# of macro Bss and mb: \# of micro BSs) with their corresponding class ratios}
	\centering
	\begin{tabular}{|c|c|c|c|}
		\hline
		Data set & Scenario & MB, mb & Class ratio $\left(\frac{\# of \mathrm{Ones}}{\# of \mathrm{Zeros}}\right)$\\
		\hline
		D1 & ---& --- & $0.53$\\
		\hline
		\hline
		\multirow{4}{*}{D2} & a& $75$, $5$ & $0.13$\\ 
		\cline{2-4}
		& b & $70$, $10$  & $0.39$\\
		\cline{2-4}      
		& c & $40$, $40$  & $2.51$\\
		\cline{2-4}      
		& d & $10$, $70$  & $19.31$\\
		\hline
		\hline
		\multirow{3}{*}{D3} & a & $50$, $10$  & $0.02$\\ 
		\cline{2-4}
		& b & $30$, $30$  & $0.15$\\
		\cline{2-4}      
		& c & $20$, $40$  & $0.26$\\
		\hline
	\end{tabular}
	\label{Tab:ClassRatioTable}
\end{table}

\subsection{Experimental setup 1}
\label{Subsec:FirstDataset}
In this experimental setup, the data-set D1 is collected from a real world deployment using massive MIMO channel sounder \cite{ArnoldMassiveMIMOChSounder}. The data-set was originally conceived for UE positioning using UL-channel impulse response, but we use the same data-set for the secondary carrier prediction. Below we describe how the measurement for the data-set is done\footnote{Recently IEEE CTW organized a data competition for estimating the $ \left(x,y,z\right) $ coordinates of UE given its UL channel impulse response \cite{IEEEData}}. 

This data set consists of an UL channel impulse response between a moving transmitter and an 8x8 antenna array (horizontally polarized patch antennas). The transmitter was moved around a residential area and transmitted UL orthogonal frequency-division multiplexing (OFDM) pilots with a bandwidth of 20 MHz and 1024 sub-carriers at a carrier frequency of 1.27 GHz. Ten percent of the subcarriers were used as guard bands, leaving 924 usable subcarriers. At the receiver 8 of the 64 antennas were perpetually malfunctioning, and hence, only 56 antennas provided useful measurements. For every location, 5 channel measurements were recorded. Figures \ref{fig:D1MapWithPath} and \ref{fig:D1PathWithSecCells} show the location of the transmitter in a local Cartesian coordinate system with the receiver placed at the origin on the XY–plane.

The UL channel measurement contains real and imaginary parts of the estimated channel matrices and the signal to noise ratio (SNR) measured at each antenna during channel estimation. The binary estimated signal denoting the existence or non-existence of the secondary carrier serve as the labels. These labels are artificially created as follows: we assume the existence of secondary cells at few locations as shown in Figure \ref{fig:D1PathWithSecCells}. In typical deployments, the secondary carriers are used to provide high speed connectivity in dense hot-spot regions. In the shown street maps, of Figure \ref{fig:D1PathWithSecCells}, these hot-spots could arise in the street intersections. So we consider locations within a circle of radius $r=0.08$ (after normalization of X and Y coordinates i.e., $X,Y \in [-1,1] \times [-1,1]$ ) and centered around secondary BSs to have a link to the secondary carrier and label them as "1". The rest of the locations are labeled as "0" and is as shown in the Figure \ref{fig:R0p08} with red (label = 1) and blue (label = 0) colored points, i.e., the label $ l_{i} $, for the BS $ b_{j} $ at location $ q_{i} $ is $ 1 $ if $ \Vert q_{i}-b_{j}\Vert_{2}\leq 0.08 $, otherwise $ 0 $
and $\Vert.\Vert_{2} $ is the usual Euclidean norm.

\subsection{Experimental setup 2}
\label{Subsec:D2 and D3}
In the previous section, we discussed the simulation setup for the synthetically created data set. Next, we describe a setup where we take  large city deployments with standard defined propagation environment \cite{CityDeployments} and generate two data sets D2 and D3. For D2 both $ f_{p} $ and $ f_{s} $ are in $ \mathrm{FR}_{1} $ and for D3, $ f_{p} \in \mathrm{FR}_{1} $ and $ f_{s} \in \mathrm{FR}_{2} $. Here we consider $1000$ UEs and 80 BSs, $f_{p}=0.9$ GHz, and $f_{s}=2$ GHz respectively (data set D2). Similarly, we consider $100000$ UEs and $95$ BSs with $f_{p}=3.5$ GHz and $f_{s}=28$ GHz respectively (data set D3). The data set in both the deployments are generated as follows: for a given carrier frequency $f_{p}$, the UL channel response for all the links above certain threshold was captured and for the same position labels are attached based on strength  of the secondary carrier link. 
We have considered different scenarios with varying number of primary (macro) and secondary (micro) BSs. These BSs are non-colocated in the sense that micro BSs are at a different geographical location than macro BS. 

Figures \ref{fig:AsianCityScenario80BS1000UE5MiBs}-\ref{fig:AsianCityScenario80BS1000UE70MiB} illustrates Asian city scenario with varying number of non co-located macro and micro BSs and similarly Figures \ref{fig:Scenario600UE50Ma10MiBs}-\ref{fig:Scenario600UE20Ma40MiBs} shows different scenarios of London city. In each of the figures green points ($\color{green}\bullet$) are the UEs having label "0", red points ($\color{red}\bullet$) are the UEs having label "1", blue stars ($\color{blue}\bigstar$) are macro BSs, and orange stars ($\color{orange}\bigstar$) are the micro BSs. From the described scenarios, we generate the data set $ \left(x,y\right) $. All the data sets D1, D2, and D3 are imbalanced in the sense that ratio of cardinality of majority to minority class labels is not one and the details are provided in Table \ref{Tab:ClassRatioTable}. We have derived few features from the channel frequency response such as energy, minimum, maximum values, distance of nearest micro BS from the macro BS from which UE is being served, cartesian coordinates of serving macro BS and the nearest micro BS and the input may contain one or more of these features along with sector ID.

In order to mitigate the effect of imbalanced data, we have used different data pre-processing techniques, such as synthetic minority oversampling technique (SMOTE) \cite{Chawlasmote}, under sampling, and hybrid sampling. In this work hybrid sampling technique is used to balance the data. Hybrid sampling involves up-sampling minority class members and down sampling majority class members in such a way that class ratio is one.

\subsection{Results}
\label{Subsec:Resuls}
The performance of binary classification experiment using random forest algorithm  with balanced input data set is analyzed using receiving operating characteristic curve (ROC). The ROC curve is obtained by plotting the true positive rate (TPR) against false positive rate (FPR) by varying classification threshold of a binary classifier and the performance is quantified using area under receiving operating characteristic curve (AUROC). The diagonal line in the ROC plot indicates random guessing similar to tossing of a fair coin experiment and has area equal to 0.5. The perfect classifier has an area equal to one. For each experiment we have performed six-fold cross validation (CV) and plotted ROC for each of the CV data sets.

The ROC curve for experimental setup 1 is shown in Figure \ref{fig:IEEEDataROC}. We see that mean AUROC is 0.93 indicating that prediction of existence or non-existence of secondary carrier using UL channel impulse response with no dedicated signaling involving UE for secondary carrier prediction. Also, from Figures \ref{fig:MAPROC} and \ref{fig:D1ROCImbalancedData} it is clear that ML algorithm outperforms the MAP classifier in terms of AUROC.

The ROC plots, with data-set D2, for experimental setup 2 are shown in Figures \ref{fig:AsianCityMic05}-\ref{fig:AsianCityMic70}. It is clear that the mean AUROC is close to one and this shows that simple features derived from UL channel impulse response can be used for secondary carrier prediction.  Similarly, the ROC plots, with data-set D3, are shown in Figures \ref{fig:LondonMi10}-\ref{fig:LondonMi30} with mean AUROC close to one for all the cases.

\section{Conclusions}
\label{Sec:Conclusions}

In this work secondary carrier prediction or the inter-frequency handover problem in a densely deployed telecommunication system is considered. The existing handover technique is reactive in nature resulting in sub-optimal energy utilization. This may lead to faster battery draining of a user equipment. In contrast to the down-link measurement based handover decision procedure, we consider up-link reference signals that are available in 5G antenna systems for beam-forming decisions for secondary carrier prediction.

 Experiments based on features derived from real world deployments and simulated data-sets showed the prediction of existence or non-existence of secondary carrier link based on up-link reference signals. The features considered are simple and can be derived with minimal computation complexity.

\bibliographystyle{IEEEtran}
\bibliography{ScpBiblioGraphy}

\end{document}